\documentstyle[preprint,aps]{revtex}

\begin{document}

\draft

\title{Wigner Function Description of the A.C.-Transport Through a Two-Dimensional Quantum Point Contact}

\author{{\bf Igor E. Aronov$^{1,2}$, Gennady P. Berman$^{1,3}$, David K. Campbell$^4$,\\ and Sergey V. Dudiy$^2$}}

\address{
%
$^1$ Theoretical Division and the CNLS, Los Alamos National Laboratory, Los Alamos, New Mexico, 87545, U.S.A.,\\
$^2$ Institute for Radiophysics and Electronics, 
National Acadamy of Sciences of Ukraine, 12 Acad. Proskura St., 310085, Kharkov, Ukraine,\\
$^3$ Kirensky Institute of Physics, 660036, Krasnoyarsk, Russia,\\
$^4$Department of Physics, University of Illinois at Urbana-Champaign,
1110 West Green St., Urbana, IL 61801-3080, U.S.A.}
\maketitle

\begin{abstract}
We have calculated the admittance of a two-dimensional quantum point contact (QPC) using a novel variant of the Wigner distribution function (WDF) formalism. In the semiclassical approximation, a Boltzman-like equation is derived for the {\it partial WDF} describing both propagating and nonpropagating electron modes in an effective potential generated by the adiabatic QPC. We show that this quantum kinetic approach leads to the well-known stepwise behavior of the real part of the admittance (the conductance) \cite{glaz1},  and of the imaginary part of the admittance (the emittance), in agreement with the latest results derived in \cite{b1}, which is determined by the number of propagating electron modes.

It is shown, that the emittance is sensitive to the geometry of the QPC, and can be controlled by the gate voltage. We established that the emittance has contributions corresponding to both quantum inductance and quantum capacitance. Stepwise oscillations in the quantum inductance are determined by the harmonic mean of the velocities for the propagating modes, whereas the quantum capacitance is a significant mesoscopic manifestation of the non-propagating (reflecting) modes. 
\newline
\renewcommand{\baselinestretch}{1.656}

\end{abstract}
\newpage 
\section{Introduction}
\label{sec-1}

        Recent technological progress in manufacturing small-scale solid
state structures has made possible the fabrication of devices involving
two-dimensional electronic systems (2DES) in the quantum
ballistic regime. One particular system that has attracted considerable
attention is the quantum point contact (QPC) (see, e.g.,  \cite{glaz1,1,2,3,4,5,6,7,9,imry,10,11,12,13,14,15,16,b1,b2,b3,b4}),
which is fabricated
by putting a split gate on the top of a $GaAs-AlGaAs$ heterostructure, thereby
creating a narrow constriction in a two-dimensional electron gas (2DEG).
Since in the ballistic regime the electrons do not experience any collisions,
through the point contact is analogous to propagation of the electromagnetic
wave through a waveguide. The width of the QPC which is controlled by the gate voltage
can be of the same order of magnitude as the Fermi
wavelength and governs the number of modes that
can propagate through the constriction.

In QPC systems several experimental investigations \cite{1,2,3,4,5,6,7} have
demonstrated quantum coherent phenomena, including quantization of
the d.c.-conductance versus the gate voltage (or the number of propagating
modes through the QPC). The theory of this phenomenon
\cite{glaz1,9,imry,10} explains the d.c.-conductance quantization as a
consequence of adiabatic transit of an electron wave through the QPC
with smooth boundaries. In an adiabatic geometry (see Fig. 1), which
is smooth on the scale of the Fermi wavelength, the longitudinal and
transverse motion of electrons can be (approximately)
separated in the Schr\"{o}dinger
equation \cite{glaz1,imry}. In this case the number of the transverse quantization modes is an adiabatic invariant, and the transverse energy plays the role of the
potential energy for the one-dimensional longitudinal motion of each mode.
Depending on whether the total energy of a given electron state is greater
or less than the effective potential energy of a given mode, the
the mode is propagating or non-propagating (see Fig. 2).

To date both  experimental and theoretical studies of QPCs have
been devoted mainly to investigations of the d.c.-conductance and the d.c.-transport.
It is clear, however, that the investigation of
the {\it a.c.-transport} can provide additional information,
since a finite frequency introduces a new time-scale and may reveal qualitatively new
effects, particularly if the new time-scale is of the order of other
characteristic times of the system.
The a.c.- conductance has been considered by M.~B\"{u}ttiker {\it et al.}
\cite{13,14,15,16,b1,b2,b3,b4}, who established that
the a.c.-transport is described by the a.c.-admittance  $Y=1/Z=G-i\omega{\cal E}$ at frequency
$\omega$, where $Z$ is the impedance. (The authors of \cite{rt,rt1,rt2,rt3} considered the a.c. - kinetic response of the resonant tunnel junction.) The
real part of the a.c.-admittance, $G$, is the conductance, and the
imaginary part of $Y$, which is proportional to ${\cal E}$,  was first introduced by
M.~B\"{u}ttiker \cite{16} as the {\it emittance}. In the important papers \cite{13,14,15,16,b1,b2,b3,b4}, the general expression for the electrochemical capacitance and for the displacement current were derived, and the steplike behavior of the QPC emittance in synchronism with the conductance steps, was established. Christen and B\"uttiker \cite{b2} also discussed the low-frequence QPC emittance of the quantized Hall conductors, and in \cite{b3} the authors used the scattering approach for the investigation of the nonlinear current--voltage characteristic of mesoscopic conductors. In papers \cite{13,14,15,16,b1,b2,b3,b4}, the emittance was expressed in terms of the geometric capacitance, transmission probability, and the densities of states of the ``mesoscopic capacitor plates'' \cite{b1}. 

In the present paper we extend the studies of the response
of a QPC to an a.c. field by developing a simple method, based on the Wigner
distribution function (WDF) formalism \cite{18,19}, for calculating
the transport characteristics. Our approach allows us to represent the emittance in terms of the capacitance and the inductance, which are expressed in the explicit form through the microscopic characteristics. The effectiveness of the WDF approach to the
modeling of small mesoscopic devices was demonstrated in Ref.
\cite{f1,f2}. In Section II, using the assumption of adiabaticity,
we derive a
Boltzmann-like quantum kinetic equation for a {\it  partial WDF} describing
transport in the quantum ballistic constriction. This equation allows us to treat the 2DES in a QPC in terms of classical trajectories for the effective 1D
motion. The electron-electron  Coulomb interaction is taken into account
within the self-consistent field approximation. In Section III we demonstrate how the a.c.-admittance of the QPC can be calculates from the propagating and non-propagating (reflected) electron
modes. Our approach recovers the quantized behavior as a function
of gate voltage of the real part of the
admittance (the conductance), consistent with previous calculations \cite{glaz1,imry} using the Landauer formula \cite{20}. 
Our approach also allows us to demonstrate that
the emittance ${\cal E}$ has the negative part, which is a quantum
inductance, to which all the propagating electron modes contribute and
whose value is determined by the harmonic mean of the electron velocities in
the quantized electron modes. The non-propagating electron modes determine the
positive contribution to the emittance ${\cal E}$, which is
a quantum capacitance, and which depends on a geometrical form of the
QPC as controlled by the gate voltage.
The conclusions are outlined in Section IV.

\section{Kinetic Equations in a Quantum Ballistic Constriction}
To find the conductivity of a 2DES in a QBC form (see
Fig. 1) taking into account both frequency dependence and spatial dispersion,
we will apply the approach based on the Wigner distribution function (WDF) \cite{18,19},

\begin{equation}
f_{\vec p}^W(\vec r)=\int d\vec r^\prime Tr\left\{\hat\rho
\exp\left[-\frac{i}{\hbar}\left(\vec p+\frac{e}{c}\vec A(\vec r)\right)
\vec r^\prime\right]\Psi^+(\vec r-\vec r^\prime/2)
\Psi(\vec r+\vec r^\prime/2)\right\}.
\label{2.1}
\end{equation}

Here $\hat\rho$ is the statistical operator of the system;
$\Psi^+(\vec r)$ and $\Psi(\vec r)$ are, respectively, the Fermi operators of creation
and annihilation of particles at the point $\vec r$; and $\vec A$
is the vector-potential of the electromagnetic field. When the
characteristic scale of the spatial inhomogeneity exceeds both the radius of interaction
among the particles and the electron's de Broglie wavelength, the kinetic
equation for the WDF (1) assumes a form equivalent to the classical
kinetic equation \cite{19},

\begin{equation}
\frac{\partial f_{\vec p}^W}{\partial t}+
\vec v\frac{\partial f_{\vec p}^W}{\partial\vec r}+
e\left\{\vec E+\frac{1}{c}\left[\vec v,\vec B\right]\right\}
\frac{\partial f_{\vec p}^W}{\partial \vec p}=
\hat I\left\{f_{\vec p}^W\right\},
\label{2.2}
\end{equation}
where as usual $\vec E$ and $\vec B$ are the electric and magnetic fields,
and $e$ is the charge and $\vec v$ the velocity of conduction
electrons. Eqn (\ref{2.2}) is valid for the extended (in the $x-y$-plane) 2DES,
when the typical scales of the inhomogeneity ($k^{-1}$, $d$) are much smaller
than a characteristic distance between the particles $n^{-1/2}$:
$k,1/d\ll n^{-1/2}$, where $k^{-1}$ is the wavelength of the electromagnetic field,
$d$ is a characteristic geometrical scale of the system, and  $n$ is the density
of the 2DEG. The characteristic distance between the particles is $\sim n^{-1/2}$
due to the weak screening in the 2DEG.

 The collision integral, $\hat
I\left\{f_{\vec p}^W\right\}$ ,in Eqn. (2), differs essentially from the classical
collision integral, since the quantum transitions included in
$\hat I\left\{f_{\vec p}^W\right\}$ reflect the character of the particle
statistics and the distinction of the WDF from the classical one \cite{19}. 
The equilibrium WDF sets the collision integral
$\hat I\left\{f_{\vec p}^W\right\}$ to zero.

        Using the definition of the WDF, we can express the charge density
$\rho$ and the current density $\vec j$, respectively, as \cite{18,19},

\begin{equation}
\rho(t,\vec r)=\frac{2e}{(2\pi\hbar)^2}\int d^2\vec p
f_{\vec p}^W(\vec r),
\label{2.3}
\end{equation}

\begin{equation}
\vec j(t,\vec r)=\frac{2e}{(2\pi\hbar)^2}\int d^2\vec p\vec v
f_{\vec p}^W(\vec r).
\label{2.4}
\end{equation}

Despite the evident analogy with the classical distribution function,
it is well known that the WDF does not have an interpretation as
the probability density, since it can take both positive and negative values,
but the integrated values shown in Eqns. (\ref{2.3}) and (\ref{2.4})
have the usual physical meanings.

For a finite system, when a 2DEG is located in a
bounded region (see Fig. 1) characterized by distances $d$ of the
order of the Fermi wavelength, the left-hand side of the kinetic equation
(2) changes its form.
Using a standard procedure \cite{18,19}, one can obtain the kinetic equation
for the WDF in the 2DES within the strip-like restricted region
$|y|<d(x)$, $d(x)=const$, which can be written in the form,

\begin{eqnarray}
&&\frac{\partial f_{\vec p}^W}{\partial t}+
\vec v\frac{\partial f_{\vec p}^W}{\partial\vec r}+e\left\{\vec E+
\frac{1}{c}[\vec v,\vec B]\right\}
\frac{\partial f_{\vec p}^W}{\partial\vec p}+
\label{2.10} \\ \cr
&&+\frac{4sgn(y)}{m\pi\hbar}\int\limits_{-\infty}^\infty dp_y^\prime
p_y^\prime\cos\left[\frac{2(p_y-p_y^\prime)}{\hbar}(d-|y|)\right]
f_{p_x,p_y^\prime}^W=\hat I\left\{f_{\vec p}^W\right\},
\nonumber
\end{eqnarray}
where $sgn(y)$ is the sign function. The integral term in the left-hand
side of Eqn. (5) arises from the transverse
quantization. The presence of this term precludes the naive
application of the classical treatment, based on the trajectories, for
the solution of the kinetic equation (5) for the WDF.

If $d(x)\not= const $, the kinetic equation for the WDF assumes an even more
complicated form. To overcome these difficulties, we will invoke
the {\it adiabatical} assumption \cite{glaz1} for the structure of
the QPC shown in Fig. 1. Explicitly, we shall assume that the
constriction is sufficiently long and smooth, the criterion
$$
d^\prime(x)\simeq d(x)/\tilde L\ll 1 
$$
is met (where $2\tilde L$ is the length
of the constriction), that the transport is adiabatic. With this assumption,
which has been discussed and analyzed in \cite{glaz1,imry},
the variables in the Schr\"{o}dinger equation can be separated,
and the eigen-wave function can be written in the form,

\begin{equation}
\psi_n(x,y)=\psi_n(x)\Phi_n[y,d(x)],
\label{2.5}
\end{equation}
where the transverse wave function

\begin{equation}
\Phi_n(y)=\frac{1}{\sqrt{d(x)}}\sin\left\{\frac{\pi
n[y+d(x)]}{2d(x)}\right\} \theta[d^2(x)-y^2],
\label{2.6}
\end{equation}
should satisfy the boundary conditions:

\begin{equation}
\Phi_n(y)\Biggr|_{y=\pm d(x)}=0,
\label{2.7}
\end{equation}
and $\theta(x)$ is the Heavisaid single-step function.

One can then derive an effective Hamiltonian for
the longitudinal wave function $\psi_n(x)$ as

\begin{equation}
\hat H=-\frac{\hbar^2}{2m}\partial_{xx}^2+\varepsilon_n(x)+e\phi(x).
\label{2.8}
\end{equation}
In (9) $\phi(x,y)$ is an electric potential, and $\phi(x)$ is the averaged
electric potential with respect to the transverse coordinate $y$,
$$
\phi(x)=\frac{1}{2d(x)}\int\limits_{-d}^d dy\phi(x,y).
$$
  The electric potential $\phi(x,y)$ is assumed to by a smoothly
varying function of the transverse coordinate $y$ within the constriction
region $|y|<d(x)$. Due to the transverse quantization, the energy
of the transverse motion $\varepsilon_n(x)$ in the Hamiltonian (9) has the form,

\begin{equation}
\varepsilon_n(x)=\frac{\pi^2n^2\hbar^2}{8md^2(x)}.
\label{2.9}
\end{equation}

With our assumptions, the transverse quantum number $n$ is
an adiabatic integral of motion. Hence
we can consider the motion of electrons in the QPC as for a set of
effective one-dimensional electron systems enumerated by
$n$. Each effective electron system is located in both
the potential $\varepsilon_n(x)$ and
the self-consistent electrical potential $\phi(x)$.
We can introduce the {\it partial WDF} (PWDF) as

\begin{equation}
f_n^W(x,p_x)=\int dx^\prime\exp\left(-\frac{ip_xx^\prime}{\hbar}\right)
Tr\hat\rho\Psi^+_n(x-x^\prime/2)\Psi_n(x+x^\prime/2).
\label{2.11}
\end{equation}

Using (11), we can represent the WDF in the form,

\begin{equation}
f_{\vec p}^W(\vec r)=\sum_{n=1}^\infty
f_n^W(x,p_x)\int\limits_{-\infty}^\infty dy^\prime
\exp\left(-\frac{ip_yy^\prime}{\hbar}\right)\Phi_{n,x}(y-y^\prime/2)
\Phi_{n,x}(y+y^\prime/2).
\label{2.12}
\end{equation}

We can derive the equation for the PWDF (11) with the use of the Wigner
transformation \cite{18,19}:

\begin{equation}
\frac{\partial f_n^W}{\partial t}+
v_x\frac{\partial f_n^W}{\partial x}+
\left[-\frac{\partial\varepsilon_n(x)}{\partial x}+eE(x)\right]
\frac{\partial f_n^W}{\partial p}(x,p)=\hat I\left\{f_{\vec p}^W\right\},
\label{2.13}
\end{equation}
where $p\equiv p_x$ and
$$
E(x)=-\frac{\partial \phi(x)}{\partial x}.
$$

In terms of the PWDF the nonequilibrium charge density and current density
can be defined as:

\begin{equation}
\rho(x,y)=\sum_{n=1}^\infty\rho_n(x)\Phi_n^2(y),
\label{2.14}
\end{equation}

\begin{equation}
j(x,y)=\sum_{n=1}^\infty j_n(x)\Phi_n^2(y),
\label{2.15}
\end{equation}
where $\rho_n(x)$ and $j_n(x)$ are the partial charge and current
densities:

\begin{equation}
\rho_n(x)=\frac{e}{\pi\hbar}\int\limits_{-\infty}^\infty dp
\left[f_n^W(x,p)-f_n^{W(0)}(x,p)\right],
\label{2.15-1}
\end{equation}

\begin{equation}
j_n(x)=\frac{e}{\pi\hbar m}\int\limits_{-\infty}^\infty dppf_n^W(x,p).
\label{2.16}
\end{equation}
In (16), $f_n^{W(0)}(x,p)$ is the equilibrium PWDF.

The motivation for introducing the PWDF is now clear. In contrast to
Eqn. (\ref{2.5}), the kinetic equation
(\ref{2.13}) describing for the PWDF {\it does} have the form of a
classical kinetic equation in the presence of an effective potential
$\varepsilon_n(x)$. Hence the solution of this equation {\it can} be
described by the characteristics, {\it i.e.}, by the classical trajectories.

        The formalism used in this paper is based on the assumption that
the kinetic equation for the WDF can also involve the collision integral.
It is well-known (see, e.g., Ref.~\cite{19}) that this can be realized at
crystal periodicity violation, which is a source of electron scattering not
distorting (or distorting weakly) the electron spectrum of the ideal
crystal. In this way a weak disorder can be taken into account within the
WDF formalism. Certainly, the impurity scattering in a form of the
collision integral for the WDF must be treated self-consistently using,
{\it e.g.}, the self-consistent Born approximation, which is the simplest method
that is free from divergences. 
In other words, the collision integral can be
described in terms of the relaxation frequency depending on the electron
energy. It is clear, that in the case when the current carriers have a high
mobility, and if the frequencies of the electromagnetic field are
sufficiently high, the approximation
for the collision integral is
justified. The forms of the electron-phonon and electron-impurity collision
integrals are too complicated \cite{19}.
However, we shall consider here the
effects associated with the linear response to the electric field. In this
case, the WDF can be found in a linear approximation with respect to the
external electric field $\vec E$. It is well-known \cite{19} that for
describing the high-frequency effects ($\omega\gg\nu$) in a sample with
a high electron mobility, the collision integral can be treated in terms of
the momentum relaxation frequency $\nu$, while the mean free path time is
$1/\nu$. 

In other words,  the collision integral in (13) includes
quantum transitions \cite{19} and intermixing of the
different electron modes (the different PWDF).
Below, we assume a quasi-ballistic regime of transport through the QPC, and will approximate the collision integral by a single momentum relaxation frequency, 

\begin{equation}
\hat I_n\left\{f_{\vec p}^W\right\}=-\nu\left[f_n^W(x,p)-f_n^{W(0)}\right],
\label{2.17}
\end{equation}
where $f_n^{W(0)}$ is the equilibrium PWDF.
The equilibrium distribution function $f_n^{W(0)}$ within the
adiabatic approximation is given by,

\begin{equation}
f_n^{W(0)}(x,p)=n_F\left\{\frac{p^2/2m+\varepsilon_n(x)-\mu}{T}\right\},
\label{2.18}
\end{equation}

$$
n_F(x)=(1+e^x)^{-1}.
$$
The function $n_F(x)$ is the Fermi function with the effective chemical potential
$\mu-\varepsilon_n(x)$, where $\mu$ is the equilibrium chemical potential
of the 2DEG. The effective chemical potential varies  smoothly as a function
of the longitudinal coordinate $x$. In this paper we are interested in the
linear response, so we expand the PWDF about its
equilibrium form
\begin{equation}
f_n^W(x,p)=f_n^{W(0)}(x,p)+f_n(x,p).
\label{2.19}
\end{equation}
The kinetic equation linearized in the electric field
$E(x,t)=E(x)\exp(-i\omega t)$, becomes

\begin{equation}
\frac{p}{m}\frac{\partial f_n}{\partial x}-
\frac{\partial\varepsilon_n(x)}{\partial x}\frac{\partial f_n}{\partial p}
+(\nu-i\omega)f_n=-eE\frac{\partial f_n^{W(0)}}{\partial p}.
\label{2.20}
\end{equation}

The natural method for solving the kinetic equation (\ref{2.20}) is the
method of characteristics. The characteristics of this equation are the
phase trajectories of a one-dimensional motion in the potential
$\varepsilon_n(x)$, which is determined from the integral of motion, {\it viz.}
the total energy $\varepsilon$:

\begin{equation}
\varepsilon=\frac{p^2}{2m}+\varepsilon_n(x)=const.
\label{2.21}
\end{equation}
We will consider a reflection symmetric QPC, {\it i.e.}, $d(x)=d(-x)$. For this case
the phase portrait is shown in Fig. 2. The heavy
lines in Fig. 2 denote the {\it separatrix},
which passes through the hyperbolic point $p=0$, $x=0$ and
separate the phase space into four regions, within which four sets of
phase trajectories exist.

The regions of propagating trajectories
($\varepsilon>\varepsilon_n(0)$) occupy the regions (see Fig. 2):
$$
1)\ \varepsilon>\varepsilon_n(0),\ p>0; \quad \mbox{and} \quad
2)\ \varepsilon>\varepsilon_n(0),\ p<0.
$$
The regions of non-propagating (reflecting) trajectories
($\varepsilon<\varepsilon_n(0)$):
$$
3)\ \varepsilon<\varepsilon_n(0),\ x>0; \quad \mbox{and} \quad
4)\ \varepsilon<\varepsilon_n(0),\ x<0.
$$
Within each region, one can find the solution of the kinetic equation for
the PWDF and derive the general formula for the partial charge $\rho_n$
and the current densities $j_n$. Here we consider the most interesting case,
when the temperature is very low ($T\rightarrow 0$, $T\ll\mu$),
so that we have a clear separation between propagating
($\varepsilon_n(0)<\mu$) and reflecting ($\varepsilon_n(0)>\mu$) channels.

For the ``open'' ({\it i.e.}, propagating) channels:

\begin{equation}
\rho_n(x)=\frac{2e^2}{h}\frac{1}{v_n(x)}\int\limits_{-L}^L
dx^\prime
E(x^\prime)sgn(x-x^\prime)\exp[i\omega^*\tau_n(x,x^\prime)sgn(x-x^\prime)],
\label{2.22}
\end{equation}

\begin{equation}
j_n(x)=\frac{2e^2}{h}\int\limits_{-L}^L
dx^\prime E(x^\prime)\exp[i\omega^*\tau_n(x,x^\prime)sgn(x-x^\prime)],
\label{2.23}
\end{equation}
where $\omega^*=\omega+i\nu$, $v_n(x)=\sqrt{(2/m)[\mu-\varepsilon_n(x)]}$, and

\begin{equation}
\tau_n(x,x^\prime)=\int\limits_{x^\prime}^x\frac{dx^{\prime\prime}}
{v_n(x^{\prime\prime})}.
\label{2.24}
\end{equation}
For the closed channels (reflecting modes) 

\begin{equation}
\rho_n(x)=\frac{2e^2}{h}\frac{sgn(x)}{v_n(x)}\int\limits_{x_n}^L
dx^\prime E(x^\prime sgn(x))\times
\label{2.25} 
\end{equation}
$$
\left\{sgn(|x|-x^\prime)
\exp\left[i\omega^*\tau_n(|x|,x^\prime)sgn(|x|-x^\prime)\right]-
\exp\left[i\omega^*(\tau_n(|x|,x_n)+\tau_n(x^\prime,x_n))\right]\right\}
\theta(|x|-x_n),
$$
\begin{eqnarray}
&&j_n(x)=\frac{2e^2}{h}\int\limits_{x_n}^L
dx^\prime E(x^\prime sgn(x))\times
\label{2.26} \\ \cr
&&\left\{
\exp\left[i\omega^*\tau_n(|x|,x^\prime)sgn(|x|-x^\prime)\right]-
\exp\left[i\omega^*(\tau_n(|x|,x_n)+\tau_n(x^\prime,x_n))\right]\right\}
\theta(|x|-x_n).
\nonumber
\end{eqnarray}
Here $x_n$ is the absolute value of the critical (turning) point, which
is determined by the condition

\begin{equation}
\varepsilon_n(x_n)=\mu.
\label{2.27}
\end{equation}

From Eqns. (23), (24) and (26), (27), it is apparent that
the transport through a QPC is described by highly
nonlocal (integral) operators. This
suggests that the charge and current densities at a given point $x$ are
influenced by the electrical field within the whole conductor.
Thus, the PWDF formalism allowed us to derive the charge and the current
densities as nonlocal operators with respect to the electric field.

\section{The admittance of the QPC}
\label{sec-3}

Our formulation of the kinetic equation for the PWDF allows us
to describe the adiabatic transport through a QPC. Using Eqns. (23), (24) and
(26), (27), we can calculate the charge and current densities in the QPC,
once the field distribution within the QPC is given. Of
particular experimental interest is the calculation of
the frequency-dependent the admittance of the QPC, the behavior
of which reveals more detailed information than any
static characteristics.

It is well-known that the static
conductance is fully specified by the potential difference (bias voltage)
between the right and left reservoirs and that the
while detailed electrical potential
profile does not influence it significantly \cite{glaz1}. This result
was derived using the Landauer formalism \cite{imry,20} when the
conductance was defined by the matrix of the transmission coefficients
of the electrons corresponding to the different propagating chennals.
We can readily show that this result also follows immediately
from our PWDF approach. In the ballistic regime, when

\begin{equation}
L\ll l,
\label{3.1}
\end{equation}
($2L$ is the distance between the reservoirs, $l$ is the mean free path), for
$\omega,\nu\rightarrow 0$, we find for the propagating modes (open channels):

\begin{equation}
j_n=\frac{2e^2}{h}V, \qquad V=\int\limits_{-L}^L dxE(x),
\label{3.2}
\end{equation}
and for the non-propagating modes (closed channels):

\begin{equation}
j_n=0.
\label{3.3}
\end{equation}

Using Eqn.~(\ref{2.15}), we obtain the for the total current flowing
through the QBC the result

\begin{equation}
I=\int\limits_{-\infty}^{\infty}dyj(y).
\label{3.4a}
\end{equation}

Hence the static conductance assumes the familiar form \cite{glaz1}:

\begin{equation}
G=\frac{I}{V}=\frac{2e^2}{h}{\cal N},
\label{3.4}
\end{equation}
where ${\cal N}$ is the number of the open channels:

\begin{equation}
{\cal N}=\left[\frac{2k_Fd(0)}{\pi}\right]; \qquad
\hbar k_F=\sqrt{2m\mu}.
\label{3.5}
\end{equation}
Here the brackets $[\cdots]$ stand for the integral part of
the enclosed expression. From these equations it is clear that the
static conductance does not depend on the details of the smooth function,
$d(x)$.

   More generally, we can use the formalism of the PWDF to
calculate the admittance at the frequency $\omega$. From formulas
(\ref{2.23}) -- (\ref{2.27}) one can see that the partial current $j_n$ is
a function of the longitudinal coordinate $x$ at $\omega\ne 0$. 
The continuity equation

\begin{equation}
div \vec j+\frac{\partial\rho}{\partial t}=0
\label{3.6}
\end{equation}
in the QPC at $\omega\ne 0$ takes a form:

\begin{equation}
\sum_{n=1}^\infty\frac{\partial}{\partial x}\left\{j_n-i\omega
\int\limits_{-L}^{x}dx^\prime\rho_n(x^\prime)\right\}=
\frac{\partial}{\partial x}\left\{I_{tot}\right\}=0,
\label{3.7}
\end{equation}
where

$$
I_{tot}=
\sum_{n=1}^\infty\left\{j_n-i\omega
\int\limits_{-L}^{x}dx^\prime\rho_n(x^\prime)\right\}.
$$
Note that the total current $I_{tot}$, which includes the current density
$\sum_{n=1}^\infty j_n(x)$ and displacement current
$-i\omega\sum_{n=1}^\infty
\int\limits_{-L}^xdx^\prime\rho_n(x^\prime)$
is independent of the longitudinal coordinate $x$. From
the form of Eqn. (\ref{3.7}) it is easy to see that the
displacement current vanishes within the left reservoir, so the total current is

\begin{equation}
I_{tot}=\sum_{n=1}^\infty j_n(-L),
\label{3.8}
\end{equation}
and the admittance can be determined as

\begin{equation}
Y=\frac{I_{tot}}{V}=\frac{1}{V}\sum_{n=1}^\infty j_n(-L).
\label{3.9}
\end{equation}
In the general case, we should determine the field $E(x)$ within the QBC from
the Maxwell equations and afterwards calculate the admittance. Here we
consider the long-wavelength approximation, in which 
\begin{equation}
v_n^*\gg\omega L_n.
\label{3.10}
\end{equation}
Here $v_n^*$ is the typical velocity for the electrons of the $n$-th
channel, $L_n$ characterizes the length of a region for each channel. For
the open (propagating) modes $L_n$ is the distance between the reservoirs
($L_n\sim 2L$) and

\begin{equation}
v_n^*=v_n(0).
\label{3.10.1}
\end{equation}
For non-propagating modes (closed channels)
modes, $L_n$ is twice the
distance between the turning point (\ref{2.27}) and the nearest reservoir
($L_n\sim 2(L-x_n)$). The typical velocity in this case is

\begin{equation}
v_n^*=\frac{2v_F}{\tilde L}\sqrt{x_n(L-x_n)},
\qquad v_F=\sqrt{\frac{2\mu}{m}},
\label{3.10.2}
\end{equation}
where $2\tilde L$ is the length of the constriction.

The condition (\ref{3.10}) means that the field
changes only slightly during the time that it takes an electron to
transit through the QPC. Hence Eqn.~(\ref{3.10}) is the condition for weak frequency
dispersion of the conductivity. To calculate the current of the propagating modes (open channels), we can approximate the velocity $v_n$ as

\begin{equation}
{v_n(x)}\simeq{v_n(0)}={v_n^*},
\label{3.11}
\end{equation}
and for the reflecting modes:

\begin{equation}
{v_n(x)}\simeq{v_n^*}\sqrt{\frac{{|x|-x_n}}{L-x_n}}.
\label{3.12}
\end{equation}
We approximate the form of the QBC (as in \cite{11}):

\begin{equation}
d(x)=d_0\exp\left[(x/\tilde L)^2\right].
\label{3.13}
\end{equation}
Using this approximation,
we get for the open channels:

\begin{equation}
j_n(-L)=\frac{2e^2}{h}\left(1+i\frac{\omega L}{v_n^*}\right)V,
\label{3.14}
\end{equation}
and for the closed channels:

\begin{equation}
j_n(-L)=-i\omega\frac{8e^2}{h}\frac{(L-x_n)}{v_n^*}
\int\limits_{x_n}^Ldx^\prime E(x^\prime)\sqrt{\frac{x^\prime-x_n}{L-x_n}}.
\label{3.15}
\end{equation}
Consistent with our choice of a reflection symmetric $d(x)$, let us
assume that the electric field inside the QPC is reflection symmetric,
$E(x)=E(-x)$.
In this case, the contribution of the open channels is
determined by the total voltage $V$ and is independent of the detailed
profile of the electrical potential inside the QPC. Thus, we can write the
admittance in the form:

\begin{equation}
Y=G-i\omega{\cal E},
\label{3.16}
\end{equation}
where $G=(2e^2/h){\cal N}$ is the static conductance. The emittance
${\cal E}$ of the QPC is given by the expression

\begin{equation}
{\cal E}=-G\frac{L}{\overline{v}^{(o)}}+
\frac{16}{3}\frac{e^2}{h}
\sum_{n={\cal N}+1}^{{\cal N}+\tilde{\cal N}}\frac{\xi_n}{v_n^*}(L-x_n).
\label{3.17}
\end{equation}
Here $\overline{v}^{(o)}$ is the harmonic mean of the velocities $v_n^*$
in the open channels~(\ref{3.11}):

\begin{equation}
\frac{1}{\overline{v}^{(o)}}=\frac{1}{{\cal N}}
\sum_{n=1}^{{\cal N}}\frac{1}{v_n^*}.
\label{3.18}
\end{equation}
The integer $\tilde{\cal N}$ determines the number of the closed channels:

\begin{equation}
\tilde{\cal N}=\left[\frac{2k_Fd_0}{\pi}\exp[(L/\tilde L)^2]
\right]-{\cal N},
\label{3.19}
\end{equation}
with $2L$ being the distance between the reservoirs. The discrete value $\xi_n$
characterizes the relative bias of voltage in the region $(x_n,L)$ filled
with the electrons of the $n^{th}$ reflecting channel:

\begin{equation}
\xi_n=\frac{3}{2}\int\limits_{x_n}^Ldx^\prime
\frac{E(x\prime)}{V}\sqrt{\frac{x^\prime-x_n}{L-x_n}}.
\label{3.19.1}
\end{equation}

>From Eqn. (\ref{3.17}) it follows immediately that the contribution of
the reflecting modes to the emittance
${\cal E}$ is positive, whereas the 
contribution of the propagating modes is negative. This observation
allows us to express our results concisely in terms
of the equivalent circuit shown in Fig. 3. The
admittance of the circuit is

\begin{equation}
Y= G-i\omega(C-\Lambda G^2/c^2),
\label{3.20}
\end{equation}
with
$$
\omega C\ll G, \qquad \omega\Lambda\ll c^2G^{-1}.
$$
The effective inductance in Eqn. (\ref{3.20}) is 
\begin{equation}
\Lambda=\frac{c^2L}{G\overline{v}^{(o)}},
\label{3.21}
\end{equation}
and the effective capacitance is

\begin{equation}
C=\frac{16}{3}\frac{e^2}{h}
\sum_{n={\cal N}+1}^{{\cal N}+\tilde{\cal N}}\frac{\xi_n}{v_n^*}(L-x_n).
\label{3.22}
\end{equation}
Note, that Eq. (52) coincides with the general expression for the emittance
derived in \cite{b1} (see Eq. (7) in \cite{b1}), where the emittance  was expressed in terms of the geometric capacitance, transmission probability, and the densities of states of the ``mesoscopic capacitor plates'' . Our description allowed us to represent the emittance in terms of the inductance (53), and the capacitance (54), which are expressed in the explicit form through the microscopic characteristics such as the harmonic mean of the velocities of the open channels (inductance), and the relative bias of voltage of the QPC $\xi_n$ (51), velocities $v_n^*$, and the values of the turning points $x_n$ (see (54)). It is easy to see, that the capacitance (54) and the inductance (53) 
demonstrate the stepwise behavior as the functions of the gate voltage. This stepwise behavior of the emittance,  as it was pointed out in \cite{b1}, is in a synchronism with the conductance steps, and is determined by the number of open (or closed) channels in the QPC. 

We can readily show what the emittance is a stepwise function of the
gate voltage. 
When the gate voltage approaches a point for which $2k_Fd/\pi$
is integer, and one more mode  opens (or closes), the inductance and
the capacitance in the expressions (\ref{3.21})  and (\ref{3.22})
increase to infinity.

In this case, the condition (\ref{3.10}) is violated, and the
contribution of these points to the admittance must be calculated separately. 
Let us analyze the asymptotic behavior of the emittance
in this case.
The approximation (\ref{3.11}),(\ref{3.12}) is justified only if for
all modes the parameter
$$
\gamma_n=(\mu-\varepsilon_n(0))/\varepsilon_n(0)
$$
is not too small. The situation when $\gamma_n$ becomes small for the  $n_0^{th}$ mode
($n_0={\cal N},{\cal N}+1$)
 means that the corresponding mode is near to the point where it transforms from propagating to non-propagating, or vice versa. 
When $|\gamma_{\cal N}|\ll 1$ (for an open channel),
we find that
in an inequality~(\ref{3.10}) and in Eq.~(\ref{3.18})
the typical velocity for $n={\cal N}$ is

\begin{equation}
 v_{\cal N}^*\simeq v_F\frac{L}{\tilde L}
\frac{\sqrt{2}}{{\rm ln}(4L^2/\tilde L^2|\gamma_{\cal N}|)}.
\label {3.22.1}
\end{equation}

 If $|\gamma_{{\cal N}+1}|\ll 1$ (for a closed channel)
then in an inequality~(\ref{3.10}) and in Eqs (\ref{3.17}), (\ref{3.21}):

\begin{equation}
v_{{\cal N}+1}^*\simeq \frac{16}{3}v_F\frac{L}{\tilde L}
\frac{\sqrt{2}}{{\rm ln}(4L^2/\tilde L^2|\gamma_{{\cal N}+1}|)}.
\label{3.22.2}
\end{equation}

Hence the contribution
of the ${\cal N}^{th}$ mode to the inductance is
\begin{equation}
 \Lambda_{{\cal N}}\simeq\frac{c^2}{G^2}\frac{2e^2}{h}\frac{\tilde
                L}{v_F\sqrt{2}} {\rm ln} \left(\frac{4L^2}{|\gamma_{\cal N}|\tilde
 L^2}\right),
 \label{3.22.3}
 \end{equation}
And the contribution of the $({\cal N}+1)^{st}$ mode to the capacitance is
 \begin{equation}
 C_{{\cal N}+1}\simeq\frac{e^2}{h}\frac{\tilde L}{v_F\sqrt{2}}
 {\rm ln} \left(\frac{4L^2}{|\gamma_{{\cal N}+1}|\tilde L^2}\right).
\label{3.22.4}
\end{equation}

If a channel opens (closes), and $\gamma_n\rightarrow 0$, there can be
a case of strong frequency and spatial dispersion. Because of this, at these points the system can not be treated in terms
of effective inductance and capacitance. (Note, that in Eqs. (55)-58),
when $\gamma_n\rightarrow 0$, the modules $|\gamma_{\cal N}|$ and $|\gamma_{{\cal N}+1}|$ should be substituted by $\sqrt{\gamma_{n}^2+(L/l)^2}$,
where $l$ is the electron's mean free path in the ballistic quantum constriction, $L/l\ll 1$.)

The emittance is described by the parameters of different
nature.  The inductance $\Lambda$ is determined by the velocities $v_n$
for the open channels, and the capacitance $C$ is mainly determined by the
distribution of the electrical field as well as by the location of
the turning points (\ref{2.27}). The mesoscopic emmitance can be
controlled by the gate voltage.

\section{Conclusions}
\label{sec-4}

We have developed a new approach, based on a partial Wigner distribution function,
to analyze electron a.c electron transport properties of a quantum point contact.
Treating the quantum ballistic constriction in the adiabatic approximation,
we derived a Boltzman-like equation for the partial Wigner
distribution function in an effective potential brought about by the quantized
transverse modes. We analyzed this equation in terms
of propagating and reflecting trajectories in the
quasiclassical approximation.

Our results establish that the a.c electron transport depends
directly on the the number of propagating and reflecting
modes and that certain features are sensitive to
the form of the distribution of the
electric field in the QPC. In particular, the real part of the admittance (the
conductance) is determined by the number of propagating electron modes,
and does not depend on the spatial distribution of the electric field
inside the QPC \cite{glaz1}.  The imaginary part of the admittance
(the emittance) exhibits stepwise oscillations
as a function of the gate
voltage and consists of two parts: the quantum inductance and
the quantum capacitance. The quantum inductance is determined by the
harmonic mean of the velocities for the propagating electron modes. The
quantum mesoscopic capacitance is specified by the reflecting
modes that are very sensitive to the geometry of the QPC.
The emmitance can be controlled by the gate voltage. Therefore, the
measurements of the admittance can be more informative than the measurements
of the static conductance.

It is important to stress that the effective quantum inductance
and capacitance, and the equivalent circuit, are concepts
valid within our linear response, low-frequency approximation.
For the high-frequency case, and when new propagating and non-prpagating modes
can appear or disappear, the frequency dispersion of the admittance is more
complicated than the linear one given by the equivalent circuit
of Eq. (\ref{3.20}). This case
must be considered using the self-consistent Maxwell equations for
the electric field in the QPC. We are presently investigating
this problem.

\acknowledgements

   We are grateful to L.I. Glazman, D.K. Ferry,  R. Akis and G.D. Doolen for fruitful discussions.
We thank the Theoretical Division and the Center for Nonlinear Studies of the Los
Alamos National Laboratory for hospitality during the completion of this
work. This research was supported in part by the Linkage Grant 93-1602 from the NATO
Special Programme Panel on Nanotechnology, by the Grant 94-02-04410 of
the Russian Fund for Basic Research, by the INTAS Grant No. 94-3862,
and by the Ukrainian Committee for Science and Technology (project No. 2.3/19
``Metal'').
%



\newpage
\begin{center}
{\bf FIGURE CAPTIONS}\\ \ \\
\end{center}
\quad\\
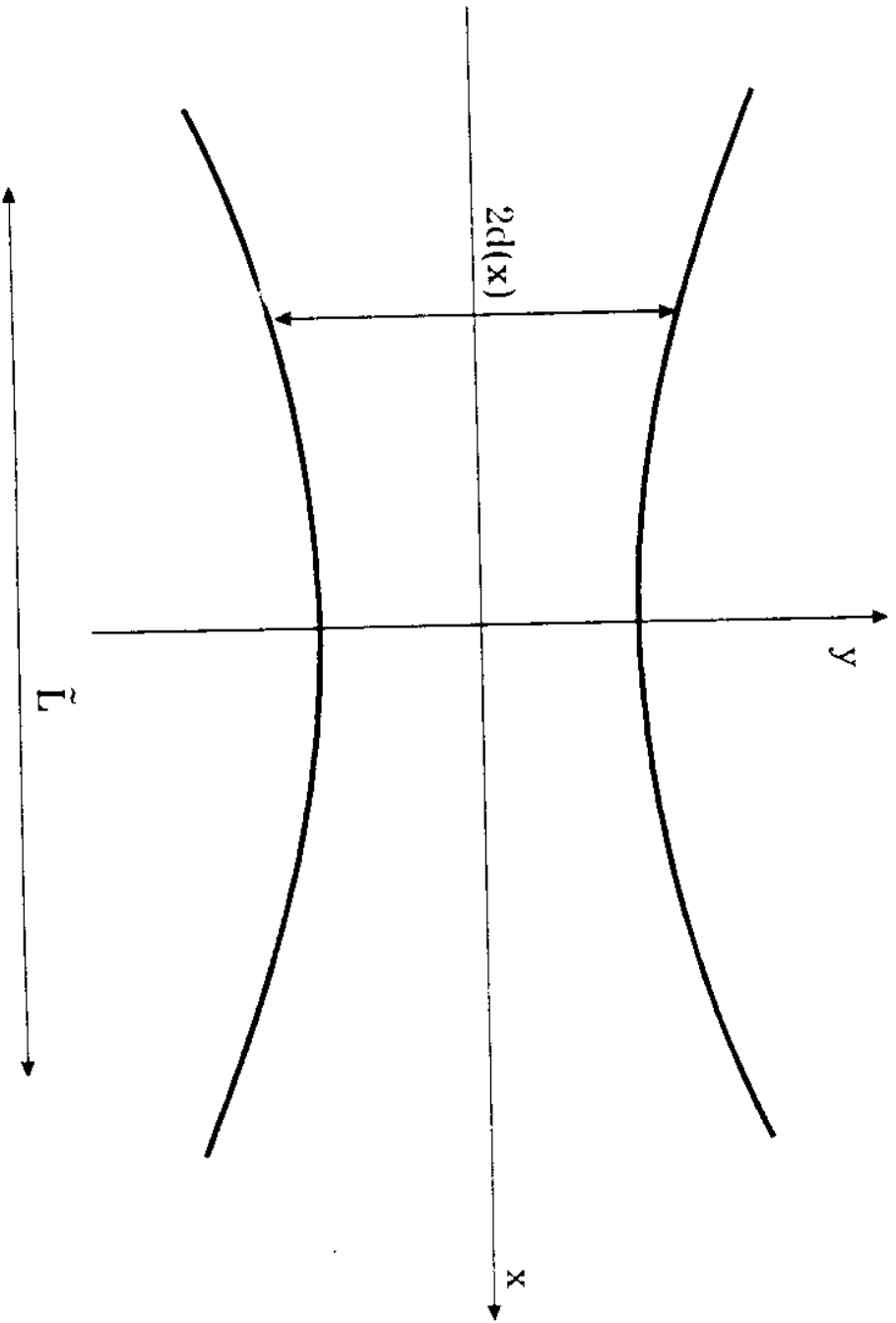
Fig. 1: {The geometry of the quantum point contact. The width is denoted by
$2d(x)$, the narrowest width is $2d_0$, and the effective length is
$2\tilde L$.}\\ \ \\
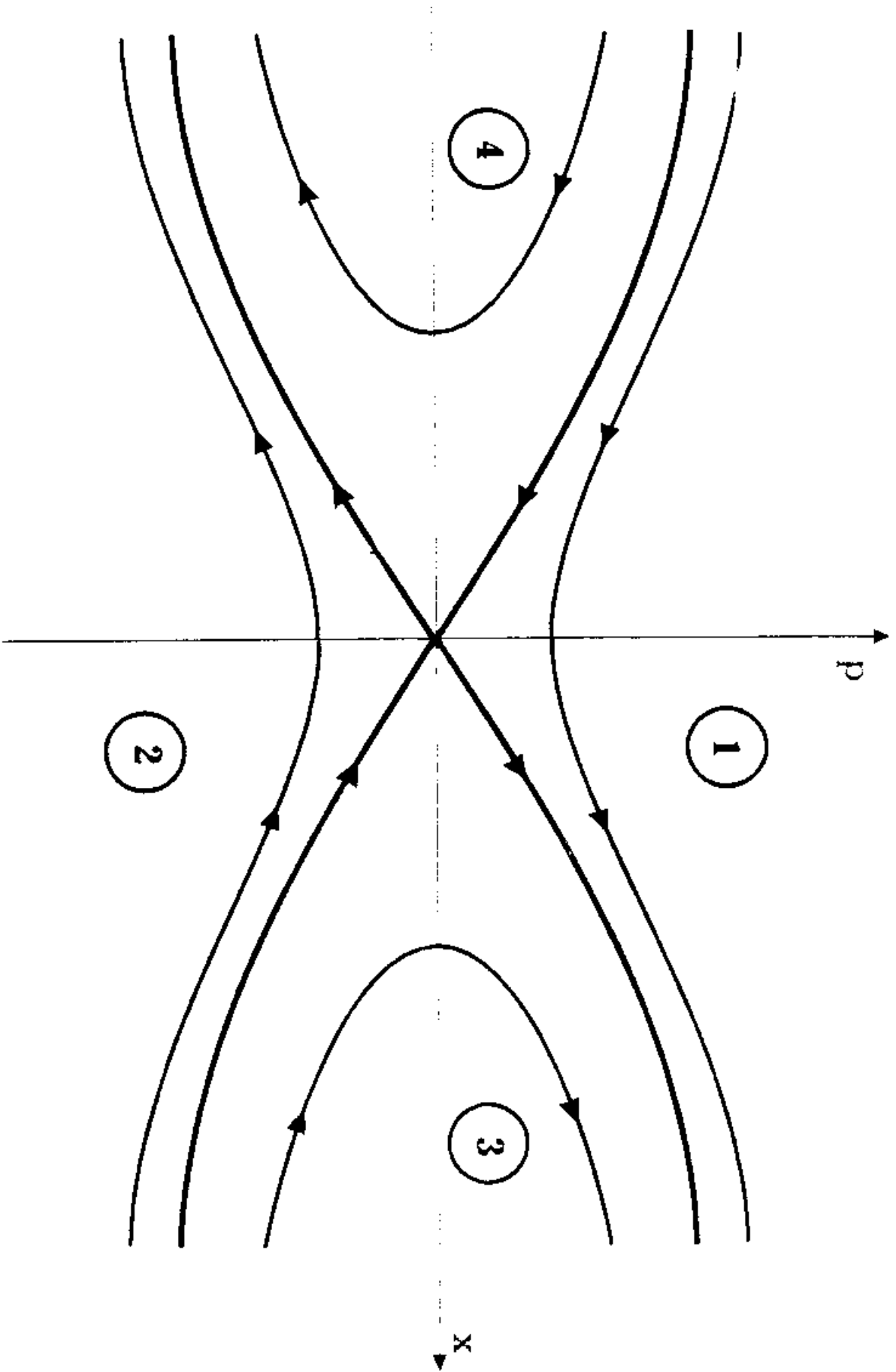
Fig. 2: {The plane of phase trajectories for one-dimensional motion
determined by the conservation of the integrals of motion. The heavy lines are
separatrices that separate the propagating modes (regions 1 and 2) and
non-propagating (reflecting) modes (regions 3 and 4)}.\\ \ \\
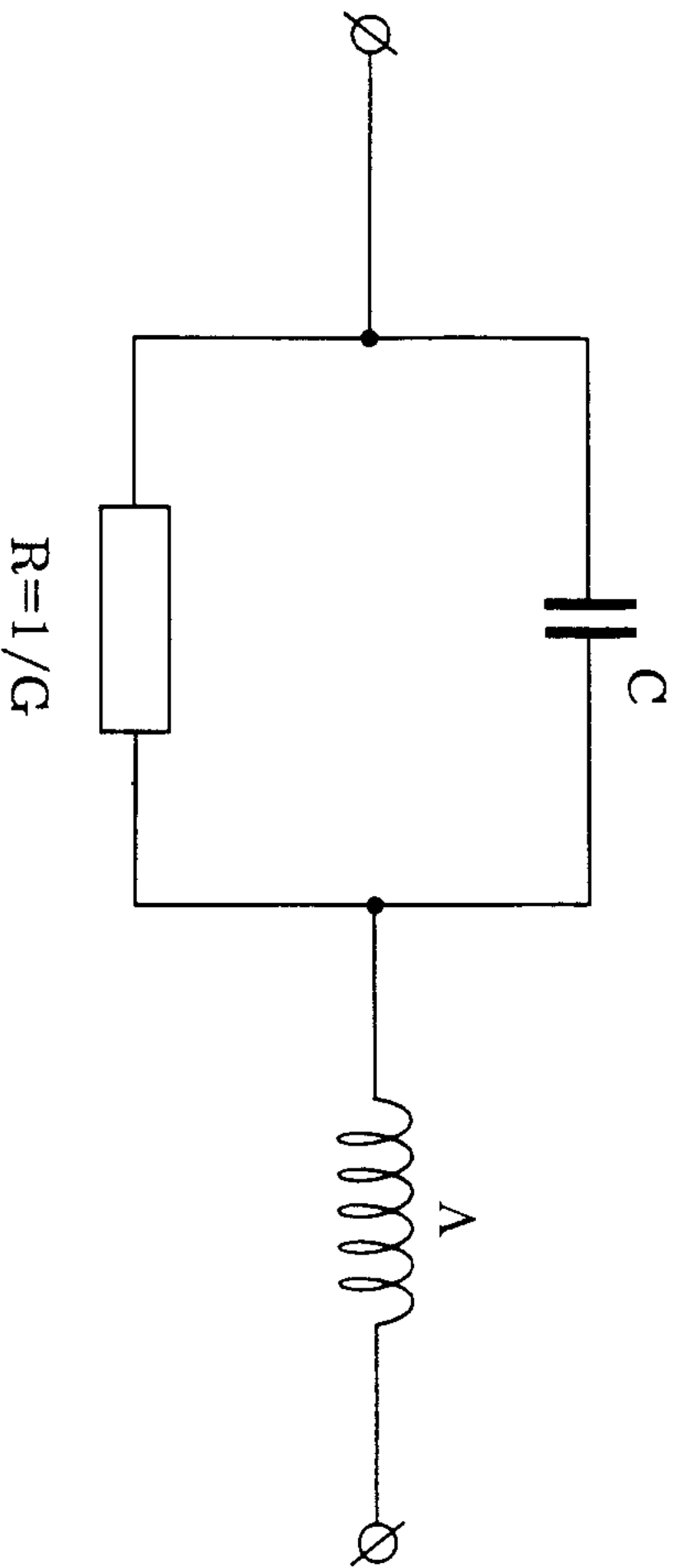
Fig. 3: {Equivalent circuit of the QPC.}


\end{document}